\documentclass[aps,prc,preprint,tightenlines,groupedaddress,showpacs]{revtex4}
\usepackage{graphicx}
\begin{document}
\preprint{DFF 394--9--02}

\title{\bf
Spinodal decomposition of expanding nuclear matter and
multifragmentation}

\author{F. Matera}
\email{matera@fi.infn.it}
\author{A. Dellafiore}
\author{G. Fabbri}
\affiliation{{\small\it Dipartimento di Fisica, Universit\`a degli 
Studi di Firenze,}}
\affiliation{
{\small\it Istituto Nazionale di Fisica Nucleare, Sezione di
Firenze,}\\
{\small\it Via G. Sansone 1, I-50019, Sesto F.no (Firenze), Italy}}


\begin{abstract}
Density fluctuations of expanding nuclear matter are studied
within a mean--field model in which fluctuations are generated by
an external stochastic field. The time evolution of the system is
studied in a kinetic--theory approach. In this model
fluctuations develop about
a mean one--body phase--space density corresponding to
a hydrodynamic motion that describes a slow expansion of the system.
A fluctuation--dissipation relation suitable for
a uniformly expanding medium is obtained and  used to constrain the
strength of the stochastic field. The coupling between the
kinetics of fluctuations and the hydrodynamic expansion
is analyzed, and the distribution of the liquid domains in the
spinodal decomposition of this expanding nuclear matter is derived. It is
found that the formation of the domains can be envisaged
as a stationary process. Comparison of the related
distribution of the fragment size with experimental
data on the nuclear multifragmentation is quite satisfactory.
\end{abstract}

\pacs{21.65.+f,24.60.Ky,25.70.Pq,21.60.Jz}
\maketitle

\section{\label{}Introduction}
In a previous paper \cite{Mat00} we have studied a model in which the
excited nuclear system that is
formed after the collision of two heavy ions at intermediate energy is
described as hot unstable nuclear matter. In this system the density 
fluctuations can lead to a liquid--gas phase 
transition ( spinodal decomposition ). The size of the liquid
domains compares well with the experimentally observed yield of
fragments. In the model of Ref. \cite{Mat00} the density is
assumed to fluctuate about a stationary mean value $\varrho_{0}$. 
In order to take into account the expansion of the complex 
system formed in the reaction, here we generalize that approach and 
consider the more realistic situation in which the density $\varrho_{0}$ 
can be time--dependent: $\varrho_{0}=\varrho_{0}(t)$.
We assume that the mean density $\varrho_{0}(t)$ changes slowly with time,
compared to the times characterizing the behavior of fluctuations,
and obtain a hydrodynamic description of the time dependence of
$\varrho_{0}(t)$. In order to describe fluctuations about this
time-dependent average behavior, we assume, like in
Ref. \cite{Mat00}, that a stochastic self--consistent mean
field induces fluctuations of the density about its average 
value. The evolution of these fluctuations is  
accordingly  described by means of a kinetic equation of mean--field 
in which a stochastic (~Langevin~) term is included.
\par
Semiclassical kinetic equations for the
one--body phase--space density have often been used for
studying the complex processes occurring in heavy
ion collisions \cite{Ber88,Schu89,Bon94}. However, in their original 
formulation, these equations account for the time
evolution of the average one--body phase--space density only, and
cannot describe phenomena  such
as multifragmentation in which
fluctuations of the phase--space density about its mean value
are believed to play an essential role
( for a review on nuclear multifragmentation see, e.g.,
Refs. \cite{Tam98,Ric01,Gup01,Bor02}~).
The Boltzmann--Langevin equation of Refs. \cite{Ay90,Ran90,Col94}
instead, incorporates also a
stochastic term into the kinetic equation and
includes fluctuations in the evolution of the phase--space density.
In Refs. \cite{Ay90,Ran90,Col94}
the diffusion coefficient of the Langevin ( fluctuating ) term is
ultimately related to the amplitude of the nucleon--nucleon scattering.
That relation is a particular case of the fluctuation--dissipation
theorem. More recently, a different method to take into account
fluctuations has been proposed in Ref. \cite{ColA98}. In that paper
the statistical fluctuations of the one--body phase--space density
are directly introduced by assuming local thermodynamic equilibrium.
A basic assumption, shared by all works 
on this subject, concerns the white--noise nature of 
the stochastic term. In Ref. \cite{Mat00} we have shown that this
assumption is compatible with the the
fluctuation--dissipation theorem only in particular situations
since this theorem provides a kind of self--consistency
relation for the stochastic field. That result is generalized here to
the case of a slowly expanding system.
\par
Thus, we extend the approach of Ref. \cite{Mat00} by considering
fluctuations about the average time--dependent density $\varrho_{0}(t)$
of nuclear matter during a hydrodynamic expansion. The evolution of the
fluctuations is described by a kinetic equation that in the 
absence of the stochastic field has a  solution corresponding to
the one--body phase--space density of the hydrodynamic motion. The
stochastic field is treated as a linear perturbation about the
zero--order hydrodynamic solution. 
The diffusion coefficient of the stochastic field is
self--consistently determined by means of a
fluctuation--dissipation relation, 
which is proved to be valid for a uniformly expanding system.
We are mainly interested in unstable situations, where
the fluctuations increase with time until they cause the
decomposition of the system. The growing of fluctuations is
essentially dominated by the unstable mean field. Thus
we focus our attention on the behavior of the mean field and
neglect the collision term in the kinetic equation.
Collisions would mainly add a damping to the growth rate
of the fluctuations and could not change the main results of our calculations,
at least at a qualitative level. 
\par
In Sec.~\ref{AA} we outline a procedure to determine the structure
of liquid domains formed within the system during its spinodal
decomposition. This formalism is a generalization of that introduced
in Ref. \cite{Mat00} and it gives a new insight into the interplay between
the hydrodynamic motion and the kinetics of fluctuations.  The present
approach allows us to determine the time behavior
of the distribution of the liquid domains, that depends on the collective
expansion energy. Moreover, our results for the expanding system
reproduce the main results of Ref. \cite {Mat00}.
\par
According to Ref. \cite{Ber83} the fragmentation phenomenon
observed in heavy ion collisions could be ascribed to a spinodal
decomposition of the bulk of nuclei. Following that suggestion, we
identify the pattern of domains formed in the decomposition of our expandig
nuclear matter with the fragments experimentally observed in
multifragmentation reactions and relate the distribution of the liquid
domains to that of the fragments.
Thus, in Sec.~\ref{BB} we compare the results of our calculations with
suitable experimental data generated by processes for which there are
indications that fragmentation could ensue from the expansion and cooling of
the emitting source, following either an initial compression
\cite{Ass99,Fran01} or an initial heating \cite{Beau99,Beau00}.
Finally in Sec. IV a brief summary and conclusions are given, while in
the Appendix the statistical properties of expanding nuclear matter
are briefly discussed.
\section{\label{AA}Formalism}
\subsection{\label{oo}Kinetic equation}
We want to study the spinodal instabilities of nuclear matter
when it is brought into the spinodal zone of the phase diagram. For
this purpose we model the system formed after the collision by a
sphere of homogeneous nuclear matter that is slowly expanding and
assume that the radius of this sphere is
much larger than the characteristic bulk lengths.
Then we can study  bulk properties of nuclear matter 
and confine our attention to the core of the sphere 
by neglecting  both surface and finite--size effects. 
We further suppose that the core of the excited system
is made of homogeneous nuclear matter with density and
temperature that are functions of time only. Hence the
continuity equation
\begin{equation}
\frac{d\varrho_0(t)}{dt}+\varrho_0(t){\bf \nabla}\cdot {\bf u}({\bf r},t)
=0\,,
\label{hydro1}
\end{equation}
requires that the divergence of the hydrodynamic velocity field
${\bf \nabla}\cdot {\bf u}({\bf r},t)$ also depends only
on time. Choosing the form 
\begin{equation}
{\bf u}({\bf r},t)=-{\bf r}\frac{\dot\gamma(t)}{\gamma(t)}
\label{vhydro}
\end{equation}
for the hydrodinamic velocity, Eq. (\ref{hydro1}) gives 
\begin{equation}
\frac{\varrho_0(t)}{\varrho_0(0)}=\frac{\gamma^3(t)}{\gamma^3(0)}
\label{dens}
\end{equation}
for the time evolution of the density. 
\par
The function $\gamma(t)$ is determined by the second Euler equation
\begin{equation}
\frac{\partial}{\partial t}{\bf u}({\bf r},t)+
\big({\bf u}({\bf r},t)\cdot{\bf \nabla}\big)\,{\bf u}({\bf r},t)=0\,,  
\label{hydro2} 
\end{equation}
which is fulfilled with    
\begin{equation}
\frac{\gamma(t)}{\gamma(0)}=\frac{1}{(1+\alpha t)}\,,
\label{gamma}
\end{equation}
where $\alpha=|\dot \gamma(0)/\gamma(0)|$ is the initial expansion rate. 
\par
Since the velocity field is simply proportional to ${\bf r}$,  
both shear and bulk viscosities do not give rise 
to any effect. Then, we can 
assume that the nuclear system moves along an isoentrope during 
the expansion. From a microscopic point of view, we shall treat
nuclear matter within a mean--field approximation. The one--body 
phase--space density appropriate to the present hydrodynamic motion is 
\begin{equation}
f_0({\bf p},{\bf r},t)=\frac{1}{2\pi^3}\frac{1}
{e^{\beta(t) \big(({\bf p}-m{\bf u}({\bf r},t))^2/(2m)-\tilde\mu(t)\big)}
+1}\,,
\label{f0}
\end{equation}
where $\beta(t) =1/T(t)$ is the inverse temperature (~we use units
such that $\hbar=~c=~k_B=1$~), and the effective chemical potential 
$\tilde\mu(t)$ is measured with respect to the uniform mean field 
$U(\varrho_0(t))$. Within this scheme the relations among  
temperature, effective chemical potential and density  
along the isoentrope are the same as for a Fermi 
gas \cite{Landau,Brack85} 
\begin{equation}
\frac{T(t)}{T(0)}=\frac{\tilde\mu (t)}{\tilde\mu (0)}=
\Big(\frac{\varrho_0(t)}{\varrho_0(0)}\Big)^{2/3}\,.
\label{isoe}
\end{equation}
\par
Thus, in our model, the average trajectory of the system in phase 
space is determined by the hydrodynamic expansion. 
Following Ref. \cite{Mat00}, we introduce now
a stochastic scalar field that generates fluctuations about the
average trajectory and calculate the density--density
response function of the system in a self--consistent mean--field
approximation. In order to derive compact analytic expressions, we use
the linearized Vlasov equation for calculating the response function.
Thus we write the one--body distribution function when
the stochastic field is present as
\[
f({\bf p},{\bf r},t)=f_{0}({\bf p},{\bf r},t)+\delta f({\bf p},{\bf r},
t)
\]
and assume that at all relevant times $\delta f<<f_{0}$. Then 
$\delta f$ obeys the equation
\begin{eqnarray}
&&\frac{\partial}{\partial t}\delta f({\bf p},{\bf r},t)+\frac{{\bf p}}{m}
\cdot{\bf \nabla}\delta f({\bf p},{\bf r},t)
\nonumber\\
&&+\frac{\partial}{\partial \tilde\mu}
f_0({\bf p},{\bf r},t)(\frac{{\bf p}}{m}-{\bf u})
\cdot{\bf \nabla}\int {\cal A}({\bf r}-{\bf r}^\prime,t)
\delta \varrho({\bf r}^\prime,t)d{\bf r}^\prime
\nonumber\\
&&
=-\frac{\partial}{\partial \tilde\mu}f_0({\bf p},{\bf r},t)
(\frac{{\bf p}}{m}-{\bf u})
\cdot{\bf \nabla}\varphi({\bf r},t),
\label{vlas}
\end{eqnarray}
where $\varphi({\bf r},t)$ is the external scalar field and 
$\delta\varrho({\bf r},t)$ is the density fluctuation 
\[
\delta\varrho({\bf r},t)\equiv
\varrho({\bf r},t)-\varrho_0(t)=\int \delta f({\bf p},{\bf r},t)d{\bf p}\,.
\]
The effective interaction ${\cal A}({\bf r}-{\bf r}^\prime,t)$
is given by the functional derivative of the mean field
\begin{equation}
{\cal A}({\bf r}-{\bf r}^\prime,t)=\frac{\delta U({\bf r},t)}
{\delta\varrho({\bf r}^\prime,t)},
\label{force}
\end{equation}
evaluated at the actual density of the expanding 
system. Here we use a schematic Skyrme--like, density--dependent,
finite--range effective interaction 
that has been introduced in Ref. \cite{ColA94}. In momentum space it reads  
\begin{equation}
{\cal A}_k=\big(a\frac{1}{\varrho_{eq}}+(\sigma+1)\frac{b}
{\varrho_{eq}^{\sigma+1}}\varrho_0^{\sigma}\big)e^{-c^2\,k^2/2}\,,
\label{inter}
\end{equation}
with 
\[
a=-356.8\,{\rm MeV},~~b=303.9\,{\rm MeV},~~\sigma=\,\frac{1}{6}\,.
\]
These values reproduce the binding energy
(~$15.75\,{\rm MeV}$~) of nuclear matter at saturation 
(~$\varrho_{eq}=0.16\,{\rm fm}^{-3}$~) and give an 
incompressibility modulus of $201\,{\rm MeV}$. 
\par
The width of the Gaussian in Eq. (\ref{inter}) has been chosen in
order to reproduce the surface-energy term as prescribed in Ref. \cite{Mye66}.
For small $k$ this interaction reduces to that employed in Ref. \cite{Mat00}
and the parameter $c$ in Eq. (\ref{inter}) is obviously related to
the parameter $d$ of Ref. \cite {Mat00}. 
\par
In the physical situations considered here
the values of temperature are sufficiently small, with respect to the
Fermi energy, so that the Pauli principle is still operating.
For $T$ sufficiently low with respect to $\tilde\mu$, the
derivative $\partial f_0({\bf p},{\bf r},t)/\partial \tilde\mu$
in Eq. (\ref{vlas}) is appreciably different from zero only
in a small domain of the nucleon velocity $|{\bf p}|/m$ about the
Fermi velocity $v_{F}$. Moreover, the expansion velocity $|{\bf u}|$ must be
smaller than $v_F$, otherwise our hydrodynamic approximation to
the expansion would not be reasonable, so we can put
 $({\bf p}/m-{\bf u})\approx {\bf p}/m$ in Eq. (\ref{vlas}).
Neglecting the expansion velocity
in Eq. (\ref{vlas}) simplifies calculations and, when taking
the space Fourier transfom,
leads to a closed equation for each Fourier coefficient.
\par
The integral equation satisfied by the response function in momentum
space reads: 
\begin{equation}
D_{k}(t,t^\prime)=D_{k}^{(0)}(t-t^\prime,t^\prime)
+{\int_{-t_0}^{t}D_{k}^{(0)}(t-t^{\prime\prime},t^{\prime\prime})
{\cal A}_k(t^{\prime\prime})D_{k}(t^{\prime\prime},t^\prime)
\,dt^{\prime\prime} }\,.
\label{resp}
\end{equation}
The origin of time is set at $t=-t_{0}$, with $t_{0}$ much larger
than the typical response times of the physical system. 
The non--interacting particle--hole propagator 
$D_{k}^{(0)}(t-t^{\prime},t^{\prime})$ has the same expression
as for a system at equilibrium, however in that case $D_{k}^{(0)}$
depends only on the time difference $t-t'$ while here it gets a
further dependence on $t'$ through the time--dependence of the
thermodynamic quantities  that determine the local--equilibrium state. 
For symmetry reasons both propagators in (\ref{resp}) depend only on 
the magnitude $k$ of the wawe vector. Equation (\ref{resp}) gives 
the response to a scalar field of a homogeneous system with 
thermodynamic properties that change in time according to
an isoentropic expansion. For a vector field, instead, the 
hydrodynamic velocity field ${\bf u}({\bf r},t)$ could 
not always be neglected, and spatial uniformity could be lacking. 
However, even for the density--density response, in our approach
the kinetic evolution of fluctuations is still coupled to
the hydrodynamic motion. 
\par
Spinodal instabilities in expanding nuclear 
matter have been previously investigated in Ref. \cite{Col95}, where
the authors focused their attention mainly on the growth of the most
unstable modes. In that paper, time--averaged values were taken for
the quantities that change on the hydrodynamic time scale, so that
the coupling between hydrodynamic motion and kinetic evolution of
fluctuations was not fully taken into account.

\subsection{\label{pp}Distribution of fluctuations} 

According to the approach of Ref. \cite{Mat00},
we assume that the time scale of terms
not included in the mean--field approximation is shorter than
the characteristic times of the mean--field dynamics and treat all
the more complicated processes like 
thermodynamic fluctuations, quantum effects and short--range
correlations, as an extra stochastic field similar to the random force
of the Langevin equation for Brownian motion. Moreover we assume that this
additional field is a Gaussian white 
noise with vanishing mean so that the time--evolution of
the density is a Markovian process. Like in Ref. \cite{Mat00}, we
shall determine the conditions
for which the white--noise assumption can be considered valid also in
this time--dependent environment.
\par
The additional stochastic mean field will induce density
fluctuations with respect to the uniform density of
the expanding system.
To be more specific, we assume that at the time $t=0$ in
the system is present a density fluctuation
$\delta\varrho({\bf r},t=0)$.
We can imagine  that this initial fluctuation is the result of
a fictitious external force that has been acting from the time $-t_0$
until time $t=0$ ( see for example Ref. \cite{Reic}, Sec. 15~I ) and
assume that density fluctuations are absent before time $-t_0$.
In linear approximation for the stochastic
mean--field, the Fourier coefficients of
$\delta\varrho({\bf r},t)$ for $t>0$ are given by
\begin{equation}
\delta\varrho_{\bf k}(t)=
\frac{\delta\varrho_{\bf k}(t=0)}
{\int_{-t_0}^{0}D_{k}(0,t^\prime )\,dt^\prime }\,
\int_{-t_0}^{0}D_{k}(t,t^\prime)\,dt^\prime
+\int_{0}^{t}D_{k}(t,t^\prime)B_{\bf k}(t^\prime)dW_{\bf
k}(t^\prime)\,.
\label{wiener}
\end{equation}
In the second integral, $B_{\bf k}(t^\prime)dW_{\bf k}(t^\prime)$  
gives the contribution of the stochastic field in the interval $dt^\prime$. 
The real and imaginary parts of the Fourier coefficients 
$W_{\bf k}(t^\prime)$ are indipendent components of a multivariate 
Wiener process \cite{Gard}. Since the stochastic 
field is real $B^{*}_{\bf k}(t)=B_{-\bf k}(t)$ and 
$W^{*}_{\bf k}(t)=W_{-\bf k}(t)$. 
\par
In Ref.  \cite{Mat00} we have shown that the
white--noise hypothesis for the stochastic field can be retained for
values of temperature and density
sufficiently close to the borders of the spinodal
region. Moreover, in the physical situations considered
both in Ref. \cite{Mat00} and here,
the strength of  particle--hole excitations
having energy larger than $kv_F$ can be neglected.
The imaginary part of
the Fourier--transformed response function
$D_{k}(\omega,t')$ displays
a sharp peak dominating the particle--hole background
at a value of $\omega\ll kv_F$. As a function of the
complex $\omega$-variable, the response function has a pole on the
imaginary axis, at a distance from the origin that is smaller
than the values of $kv_F$. The position of this pole determines
the time scale characteristic of the response function. This time 
is much longer than the times $1/(kv_{F})$ characteristic 
of the non interacting propagator 
$D_{k}^{(0)}(t-t^\prime,t^\prime)$. Thus, when taking the
inverse Fourier transform, we can neglect the contributions from the
cut along the real-$\omega$ axis and keep only that of the isolated
pole. Moreover, here we are interested in density fluctuations that
either relax towards their equilibrium values, or grow indefinitely in
the unstable case. Thus we look for solutions of Eq. (\ref{resp}) with
values of $(t-t')$ of the same order of magnitude as the damping or
growth time scales. In this case we can neglect the
inhomogeneous term in Eq. (\ref{resp}). 
\par
When our combined hydrodynamic and kinetic approach is justified,
the time typical of the hydrodynamic expansion $|\gamma(t)/\dot
\gamma(t)|$ is also longer than $1/(kv_F)$, as a consequence the
integral equation (\ref{resp}) is reduced to a simpler differential
equation.
Since the propagator $D_{k}^{(0)}(t-t'',t'')$ in the integral in
(\ref{resp}) depends on the second argument only through the slowly
changing hydrodynamic quantities, we expand it as
\begin{equation}
D_{k}^{(0)}(t-t^{\prime\prime},t^{\prime\prime})\simeq\,
D_{k}^{(0)}(t-t^{\prime\prime},t)+(t-t^{\prime\prime})
\frac{\partial}{\partial t}D_{k}^{(0)}(t-t^{\prime\prime},t)\,,
\label{resp0}
\end{equation}
and neglect higher order terms. Thus we only need the two quantities
\[
D_{k}^{(0)}(\omega=0,t)=-\frac{\partial \varrho_0(t)}{\partial \tilde\mu}
\] 
and 
\[
i\frac{\partial}{\partial \omega}D_{k}^{(0)}(\omega,t)|_{\omega=0}=
\,\frac{1}{\pi}\,\frac{m^2}{k}F(\beta \tilde\mu)\, ,
\]
where the function 
\[F(\beta \tilde\mu)=\,\frac{1}{e^{-\beta \tilde\mu}+1}\]
is constant during the isoentropic expansion, 
because of the relation (\ref{isoe}).   
\par
With these replacements, the integral equation (\ref{resp}) turns
into the following differential equation
\begin{widetext}
\begin{equation}
\frac{1}{\pi}\,\frac{m^2}{k}F(\beta \tilde\mu){\cal A}_k(t)
\frac{\partial}{\partial t}D_{k}(t,t^\prime)=
\bigg(1+\frac{\partial \varrho_0(t)}{\partial \tilde\mu}\,
{\cal A}_k(t)
-\frac{1}{\pi}\,\frac{m^2}{k}F(\beta \tilde\mu)
\frac{\partial{\cal A}_k(t)}{\partial t}\bigg)D_{k}(t,t^\prime)\,.
\label{diffresp}
\end{equation}
\end{widetext}
The solution of this equation can be put in the form
\begin{equation}
D_{k}(t,t^\prime)=C_k(t^\prime)\frac{{\cal A}_k(t^\prime)}
{{\cal A}_k(t)}e^{\int_{t^\prime}^{t}\Gamma_k(t^{\prime\prime})
\,dt^{\prime\prime}}\,, 
\label{solut}
\end{equation}
where the function 
\begin{equation}
\Gamma_k(t)=\frac{\pi}{m^2}\frac{1}{F(\beta \tilde\mu)} 
\frac{\partial\varrho_0(t)}
{\partial\tilde\mu}\,\frac{\bigg({\displaystyle \frac{\partial^2 f(t)}
{\partial\varrho_0^2}}|_T+{\cal A}_k(t)-{\cal A}_0(t)\bigg)}
{{\cal A}_k(t)}k 
\label{rate}
\end{equation}
is the damping or growing rate ( depending on its sign ) of the density
fluctuations. This function extends the result given
in Eq. (2.16) of Ref. \cite{Mat00} to a slowly changing system.
The quantity $f(t)$ in this equation is the free-energy
density.
The sign of $\Gamma_k(t)$ is negative for ${\displaystyle 
\frac{\partial^2 f(t)}
{\partial\varrho_0^{2}}|_{T}>0}$, corresponding to a system at
equilibrium, while it is positive for
\[
\frac{\partial^2 f(t)}
{\partial\varrho_0^{2}}|_T<-({\cal A}_k(t)-{\cal A}_0(t))\leq 0\,,
\]
when the system is inside the spinodal region. 
\par
In order to derive Eq. (\ref{rate}) we have used the following relation
\begin{equation}
\frac{\partial\tilde\mu}{\partial\varrho_0}|_T=\,
\frac{\partial^2 f}{\partial\varrho_0^2}|_T-{\cal A}_0\, .
\label{chem}
\end{equation}
\par
The function $C_k(t^\prime)$ in Eq. (\ref{solut}) is still to be
determined, for the moment we assume that it varies
according to the hydrodynamic time--scale and shall verify
{\it a posteriori} the validity of this conjecture.
\par
If the approximate solution (\ref{solut}) for $D_k(t,t^\prime)$
is replaced into Eq. (\ref{wiener}), the following expression 
for the density fluctuations $\delta\varrho_{\bf k}(t)$
is obtained:
\begin{eqnarray}
\delta\varrho_{\bf k}(t)=&&\delta\varrho_{\bf k}(t=0)
\frac{{\cal A}_k(0)}{{\cal A}_k(t)}e^{\int_0^t\Gamma_k(t^{\prime\prime})
\,dt^{\prime\prime}}\nonumber\\
&&+\frac{1}{{\cal A}_k(t)}\int^t_0e^{\int_{t^\prime}^{t}
\Gamma_k(t^{\prime\prime})\,dt^{\prime\prime}}
{\cal A}_k(t^\prime)C_k(t^\prime)
B_{\bf k}(t^\prime)\,dW_{\bf k}(t^\prime)\, .
\label{ornul}
\end{eqnarray}
This expression  can be viewed as  a solution of the stochastic   
differential equation
\begin{equation}
dz_{\bf k}(t)= \Gamma_k(t)z_{\bf k}(t)dt+{\cal A}_k(t)C_k(t)
B_{\bf k}(t)dW_{\bf k}(t)\,,
\label{difforn}
\end{equation} 
with $z_{\bf k}(t)=\delta\varrho_{\bf k}(t){\cal A}_k(t)$.  
The corresponding Fokker--Planck equation
for the probability distribution $P[z_{\bf k}(t)]$ reads
\begin{eqnarray}
&&\frac{\partial}{\partial t}P[z_{\bf k}(t)]=|\Gamma_k|
\frac{\partial}{\partial z_{\bf k}(t)}z_{\bf k}(t)
P[z_{\bf k}(t)]\nonumber\\
&&+\frac{1}{2}\,{\cal A}_k(t)^2C_k(t)^2 
| B_{\bf k}(t)|^2\frac{\partial^2}{\partial z_{\bf k}^2(t)}
P[z_{\bf k}(t)]\,.
\label{fpe}
\end{eqnarray}
\par
The stochastic part of the mean field is completely determined
once the coefficients $B_{\bf k}(t)$ are known. In order to obtain
information about these coefficients we concentrate our attention on
the correlations of density fluctuations at local equilibrium.
Due to the independence of the components of the multivariate
Wiener process $W_{\bf k}(t)$, only the correlation terms with
${\bf k}^\prime=-{\bf k}$ survive. The equation for the equilibrium 
fluctuations can be obtained from Eq. (\ref{wiener}) by shifting the
initial time $t=0$ to $-t_{0}$, when the stochastic field has been
turned on, without including any particular condition at later times.
Then, the correlations are given by
\begin{eqnarray}
<\delta\varrho_{\bf k}(t)\delta\varrho_{-\bf k}(t^\prime)>&&=
\frac{1}{{\cal A}_k(t){\cal A}_k(t^\prime)}\nonumber\\
&&\times\int_{-t_0}^{min(t,t^\prime)}
e^{\int_{t_1}^{t^\prime}\Gamma_k(t^{\prime\prime})\,dt^{\prime\prime}}
e^{\int_{t_1}^{t}\Gamma_k(t^{\prime\prime})\,dt^{\prime\prime}}
{\cal A}_k(t_1)^2C_k(t_1)^2|B_{\bf k}(t_1)|^2dt_{1}\, ,
\label{varia1}
\end{eqnarray}
where the brackets denote ensemble averaging. 
\par
The aim of our approach is that of determining the Fourier
coefficients $B_{\bf k}(t)$ in Eq. (\ref{wiener})
as functions of $\varrho_0(t)$ and $T(t)$ for the
system at local equilibrium, and of extending afterwards the
expressions so found to non--equilibrium cases. Such a procedure
is usually followed when treating instabilities by making use of
the fluctuation--dissipation theorem, see e.g.
Refs. \cite{Hoff95,Gunt83}. As a part of this program, we have to verify
the self--consistency of the withe--noise hypothesis for the
stochastic field, i.e. the coefficients $B_{\bf k}(t)$ must be
local functions of time. 
This step can be accomplished by exploiting the
fluctuation--dissipation relation obtained
in the Appendix within the same approximations used in deriving
Eq. (\ref{resp}). The fluctuation--dissipation relation
involves the Fourier transforms of the density correlator
and of the response function with respect to the
difference of times, whereas both the average value of the density 
correlations and the response function refer to a state that changes with 
time according to the hydrodynamic time scale $|\gamma(t)/\dot\gamma(t)|$.
The condition of local or instantaneous equilibrium
for the hydrodynamic expansion
requires that the kinetic time scales, $1/|\Gamma_k(t)|$ and
$1/v_Fk$, are smaller than $|\gamma(t)/\dot\gamma(t)|$. Thus,
in Eq. (\ref{varia1}) the quantities ${\cal A}_k(t)$, $C_k(t)$,
$|B_{\bf k}(t)|$ and $\Gamma_k(t)$ are slow functions of
time, compared to the exponentials, and can be approximated by the
value that they take  at the upper integration extreme.
Within this approximation Eq. (\ref{varia1}) gives
( choosing $t>t^\prime$ )
\begin{equation}
<\delta\varrho_{\bf k}(t)\delta\varrho_{-\bf k}(t^\prime)>=
\frac{{\cal A}_k(t)}{{\cal A}_k(t^\prime)}
C_k(t^\prime)^2|B_{\bf k}(t^\prime)|^2\frac{
e^{\Gamma_k(t^{\prime})(t-t^\prime)}}{2\,|\Gamma_k(t^\prime)|}\,.
\label{varia2}
\end{equation}
\par
In Ref. \cite{Mat00} we have shown that,
if the classical limit $\omega/T\ll 1$ (~or $|\Gamma_k(t)|/T\ll 1$~)
can be taken when evaluating terms on both
sides of the fluctuation--dissipation relation, the
assumption of a white--noise stochastic field
can be retained. Since the relevant
values of the wave vector $k$ turn out to be such that
the quantity  $kv_F$ is of the same order of magnitude as $T$,
the limit $\omega/T\ll 1$ also implies $\omega/kv_F\ll 1$.
In analogy with Ref. \cite{Mat00}, here we also assume that
$T>|\Gamma_k(t)|$, so that the result obtained there, concerning the
validity of this inequality when temperature and density are
in the proximity of the spinodal
region, still applies. Then the fluctuation--dissipation
relation of Eq. (\ref{fdr}) can be put in the form
\begin{equation}
\frac{\partial}{\partial t}
<\delta\varrho_{\bf k}(t)\delta\varrho_{-\bf k}(t^\prime)>=
-T(t^\prime)D_k(t,t^\prime)\,.
\label{fdt}
\end{equation}
\par 
The variance of the density fluctuations of nuclear matter at local 
equilibrium can be determined by using Eq. (\ref{eqtfluct}). Within the 
present approach ( see also Ref. \cite{Mat00} ) it is given by 
\begin{equation}
\sigma_k^2(t)|_{eq}=<\delta\varrho_{\bf k}(t)
\delta\varrho_{-\bf k}(t^\prime)>
=\frac{T(t)}
{f^{\prime\prime}(t)+\big({\cal A}_k(t)-{\cal A}_0(t)\big)}\,,
\label{eqfluct}
\end{equation}
where we have introduced the abbreviation
\[f^{\prime\prime}=\,\frac{\partial^2 f}{\partial\varrho_0^2}|_T\, .\]
By exploiting this equation and Eq. (\ref{fdt}),
we can obtain the following relation between the function $C_k(t)$ and
the diffusion coefficient $|B_{\bf k}(t)|$
\begin{equation}
C_k(t)=2\frac{T(t)}{|B_{\bf k}(t)|^2}
\label{funct}
\end{equation}
and the additional relation 
\begin{equation}
|B_{\bf k}(t)|^2=\frac{2}{\pi}m^2T(t)F(\beta \tilde \mu){\cal A}_k(t)
\big({\cal A}_0(t)-f^{\prime\prime}(t)\big)\frac{1}{k}\,.
\label{difcoef}
\end{equation}
These equations confirm that both $|B_{\bf k}(t)|$ and $C_k(t)$ change in
time according to the hydrodynamic time--scale, as we have
previoulsly assumed. 
\par
Both the diffusion coefficients of Eq. (\ref{difcoef}) and the function
of Eq. (\ref{funct}) have been derived for a system at local
equilibrium using the subsidiary conditions
$\beta<1/|\Gamma_k|<|\gamma/\dot \gamma|$. The second constraint
limits our approach to slow expansions
of nuclear matter. Following Refs. \cite{Hoff95,Gunt83}
( see also the discussion in Ref. \cite{Mat00} on this point ),
we assume that the relations
(\ref{funct}) and (\ref{difcoef}) for $C_k(t)$ and $|B_{\bf k}(t)|$ are
valid also in unstable situations, provided that the conditions
$\beta<1/|\Gamma_k|$ and  $|\Gamma_k|<v_Fk$
are still fulfilled, at least for time--averaged values.
\par
The solution of the Fokker--Planck equation (\ref{fpe}) 
is given by a Gaussian distribution. For simplicity we
assume that the state of the system at $t=0$ is homogeneous on the 
average (~$<\delta\varrho_{\bf k}(t=0)>=0$ for $k\not=0$~), in case
of necessity a nonvanishing mean value could easily be introduced, then
Eq. (\ref{ornul}) implies that this property holds also during the time
evolution. By making the change of variables
$\delta\varrho_{\bf k}(t)=z_{\bf k}(t)/{\cal A}_k(t)$,
we obtain for the probability distribution of density
fluctuations the explicit expression
\begin{equation}
P[\delta\varrho_{\bf k}(t)]=\,N_{1}(t)\exp
\Big(-\frac{1}{2}\sum_{\bf k}
\delta\varrho_{\bf k}^*(t)\frac{1}{\sigma^2_k(t)}
\delta\varrho_{\bf k}(t)\Big)\, ,
\label{gauss}
\end{equation}
with the variance $\sigma^2_k(t)$ given by
\begin{equation}
\sigma^2_k(t)=\sigma^2_k(t=0)e^{2\int_0^t\Gamma_k(t^\prime)
\,dt^\prime}+
\frac{2\pi}{m^2}\frac{k}{F(\beta\tilde \mu)}\frac{1}{{\cal A}_k^2(t)}
\int_0^te^{2\int_{t^\prime}^t\Gamma_k(t^{\prime\prime})
\,dt^{\prime\prime}}T(t^\prime)\frac{{\cal A}_k(t^\prime)}
{(f^{\prime\prime}(t^\prime)-{\cal A}_0(t^\prime))}\,dt^\prime\,,
\label{variance}
\end{equation}
while $N_{1}(t)$ in Eq. (\ref{gauss}) is a time--dependent
normalization factor.
\par
Equations (\ref{gauss}) and (\ref{variance}) describe
density fluctuations both in situations of local equilibrium
and for systems out of equilibrium.
In the first case, for $t\gg 1/|\Gamma_k|$ the variance
relaxes towards the typical value of equilibrium
thermodynamic fluctuations, with values of density and temperature
appropriate to that instant, ( see Eq. (\ref{eqfluct}) ), while in the
second case the variance grows exponentially for  fluctuations
with wave number smaller than a certain time--dependent value.
\par
Starting from the probability distribution for density
fluctuations given by Eq. (\ref{gauss}), we can determine the
corresponding distribution for the size of the correlation domains.
We would like to recall again that the stable and unstable cases
can be treated within the same scheme, however here we
investigate only the case in which the system has been
quenched inside the spinodal zone. Then, it develops
density fluctuations that grow with time and will
eventually lead to the decomposition.
\par
The procedure followed in Ref. \cite{Mat00} for nuclear matter  
at constant density and temperature can
directly be extended to the present expanding system,
this is justified essentially by the Gaussian form of
the probability distribution for density fluctuations.
Then the distribution of the domain size is given by
\begin{equation}
P(A,t)=\,\frac{1}{3}\frac{1}{\sqrt{\pi}}\sqrt{\frac{r_0(t)}{L(t)}}\,
A^{-5/6}e^{ -[r_0(t)/L(t)]A^{1/3}}\,,
\label{distra}
\end{equation}
where $A$ is the number of nucleons contained in a correlation
domain, $L(t)$ is the length scale of the domain pattern and
$r_0(t)$ is the mean interparticle spacing. Like in Ref. \cite{Mat00},
in order to take into account that $A$ is a discrete variable
we express the probability of finding a correlation domain
containing $A$ nucleons, $Y(A,t)$, through the integral
\begin{equation}
Y(A,t)=\,\int _{A-1}^{A}dA\,P(A,t)\,. 
\label{proba}
\end{equation}
For large $A$, $Y(A,t)$ tends to coincide with $P(A,t)$.
\section{\label{BB}Results}
\par
A reasonable assumption for the initial condition is that the
system can be modeled as nuclear matter at local equilibrium. 
This implies  that the density fluctuations are those of the 
hydrodynamic regime. In particular the variance of fluctuations 
corresponds to an equilibrium state of given density and 
temperature at that instant, with $\sigma^2_k(t=0)|_{eq}$ given by
Eq. (\ref{eqfluct}).
Values of parameters $T(0)=6\,{\rm MeV}$ and $\varrho(0)=0.7\varrho_{eq}$
are compatible with
the nuclear multifragmentation process. We also choose
the initial expansion rate $\alpha =|\dot \gamma(0)/\gamma(0)|$
in the range $((1/6)10^{-2}\div 10^{-2})~c/{\rm fm}$, which
corresponds to a gentle expansion and 
cooling of the compound system formed after 
the collision. For a sphere of nuclear matter 
of radius $R\simeq  10\,{\rm fm}$, the peripheral velocity lies in the
interval $(0.017\div0.1)c$.  Then we can investigate the
transition from a quasistationary regime to a situation where the
coupling between kinetic and hydrodynamic motions can give
non--trivial effects.
\par
The time evolution of the fluctuations is ruled essentially by 
the growth rate $\Gamma_k(t)$. This function  depends on time 
through the product $\alpha\, t$, as a consequence the time average, 
$\Gamma_{k}^{av}(t)$, scales with $\alpha$. The magnitude of     
$\Gamma_{k}^{av}(t)$ determines  the range of validity of our 
approximations. The diffusion coefficients (\ref{difcoef}) of the 
stochastic equation have been obtained in the hypothesis that  
thermodynamic fluctuations are the main effect, this requires 
$\Gamma_{k}^{av}(t)< T(t)$ and $\Gamma_{k}^{av}(t)<kv_F$. 
\begin{figure}
\includegraphics{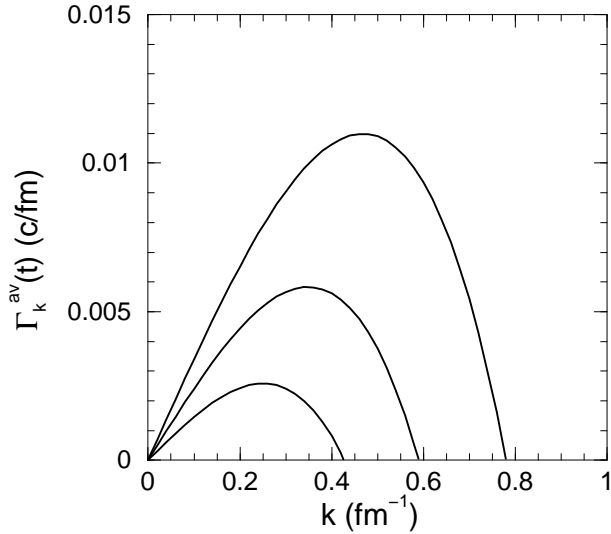}
\caption{\label{fig1}Time averaged growth--rate as a function of $k$ at 
different times, from bottom to top $t=10,30,50~{\rm fm}/c$. The value of 
$\alpha$ is $\alpha=0.01$}
\end{figure}
In Fig.~\ref{fig1} we report the mean growth rate as a function of
$k$ at three different times for a given value of the parameter
$\alpha$ ( $\alpha=0.01$ ). From that figure we can see that, for this
value of $\alpha$, the  condition $\Gamma_{k}^{av}(t)< T(t)$
is satisfied up to a time $t\lesssim 50~{\rm fm}/c$,
moreover, the relevant values of the wave
vector $k$ and the values taken by
the nuclear density in the interval $0< t < 50\;{\rm fm}/c$
are such that at each instant the quantity $kv_F$ is
less than $T$. Thus the condition $\Gamma_{k}^{av}(t)< T(t)$
also implies $\Gamma_{k}^{av}(t)< kv_F$.
According to the scaling law for the function
$\Gamma_k^{av}(t)$,
these limits are respected for longer times if $\alpha$ is smaller.
\par
\begin{figure}
\includegraphics{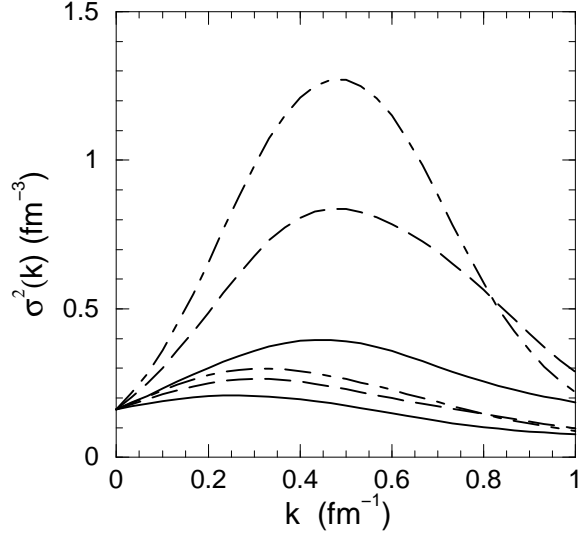}
\caption{\label{fig2}The variance as a function of $k$ for  
three different values of $\alpha$: solid lines, $\alpha=10^{-2}\,c/{\rm fm}$ 
and $t=30,50\,{\rm fm}/c$;  
dashed lines, $\alpha=(1/3)10^{-2}\,c/{\rm fm}$ and $t=60,100\,{\rm fm}/c$; 
dot--dashed line, $\alpha=(1/6)10^{-2}\,c/{\rm fm}$ and $t=80,150\,
{\rm fm}/c$. 
The upper line corresponds to the later time. 
}
\end{figure}

In Fig.~\ref{fig2} the variance $\sigma^2_k(t)$
is displayed for three different values of the parameter $\alpha$ and
at two different times for each value of $\alpha$. The times have been
chosen in such a way that the points representing the system on the
phase diagram lie cose to each other inside the spinodal region in
all three cases. The values of
density and temperature at a given instant are lower
for higher values of $\alpha$. From Fig.~\ref{fig2} we can see that
for slower expansions the variance is a faster function of time.
Thus, at the same point inside the spinodal region, the density
fluctuations are much larger for lower values of $\alpha$.
The slower the motion along an isoentrope, the easier is
the decomposition of our nuclear system.
\par
\par
The probability $Y(A,t)$ of Eq. (\ref{proba}) is completely
determined once the ratio between
the length scale $L(t)$ and the mean interparticle spacing $r_0(t)$
has been fixed. The parameter $L(t)$ is the length characterizing
the decrease of the correlation function with distance, while the
correlation function itself is the Fourier transform  
of the variance $\sigma^2_k(t)$ with respect to $\bf{k}$. 
A sufficiently accurate expression of the length $L(t)$ is
given by the simple relation
\begin{equation}
L(t)=\,\frac{\pi}
{k_M(t)+\Delta_k(t)}\,, 
\label{correl}
\end{equation}
where $k_M(t)$ is the value of $k$ at the maximum of $\sigma^2_k(t)$
and $\Delta_k(t)$ is the width of the broad peak in
Fig.~\ref{fig2}. This simple estimate of $L(t)$
is in good agreement with the more precise evaluation made in Ref.
\cite{Mat00}
for the stationary case. By comparing Figs.~\ref{fig1} and
\ref{fig2} it can be seen that the positions of
the maximum of $\Gamma_{k}^{av}(t)$ and of $\sigma^2_k(t)$
tend to coincide with increasing $t$.
\par
\begin{figure}
\includegraphics{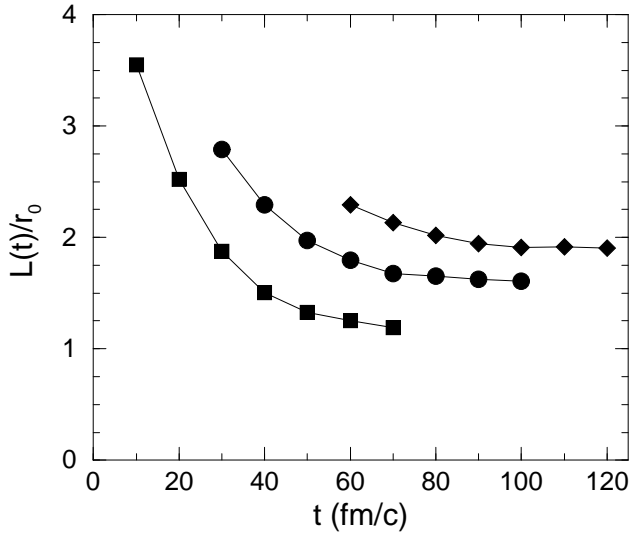}
\caption{\label{fig3}Behavior of the ratio between the length scale 
and the mean interparticle spacing $L(t)/r_0(t)$ for  
three different values of $\alpha$: squares, $\alpha=10^{-2}\,c/{\rm fm}$; 
circles, $\alpha=(1/3)10^{-2}\,c/{\rm fm}$; 
diamonds, $\alpha=(1/6)10^{-2}\,c/{\rm fm}$ . 
}
\end{figure}

In Fig.~\ref{fig3} the ratio $L(t)/r_0(t)$ is shown as a function of
time for three different values of the expansion rate $\alpha$. The
time interval for each of the three cases has been chosen so
to allow for a comparison  of the ratio $L(t)/r_0(t)$
for the same values of $\varrho(t)$ and $T(t)$. In all cases
the ratio $L(t)/r_0(t)$ approaches an almost constant value,
lower than the initial one, for sufficiently long times. This fact can be
ascribed to two competitive effects: with increasing $t$
the broad peak of $\sigma^2_k(t)$ shown in Fig.~\ref{fig2} gets 
narrower, at the same time 
$k_M(t)$ increases because the system reaches deeper
regions inside the spinodal zone. The steeper initial slope seen in
Fig.~\ref{fig3} instead, can be 
attributed to the fact that both $k_M(t)$ and $\Delta_k(t)$
increase with time, while $\sigma^2_k(t)$ changes from its
initial value. This process is faster for rapid expansions. 
The ratio $L(t)/r_0(t)$ is the only  parameter
contained in the probability distribution
$Y(A,t)$. The formation of the pattern 
of correlation domains  can be viewed as a stationary process
starting from the beginning of the plateau of $L(t)/r_0(t)$,
even if the nuclear system is still expanding.
In Fig.~\ref{fig3} the last point (~$t=70\,{\rm fm}/c$~)
of the plot with $\alpha=0.01\,c/{\rm fm}$ should be taken with
caution since it lies beyond the estimated range of validity of
our calculations. 
\par 
In the present scenario, once expansion   
and cooling have brought the bulk of the system
within the spinodal zone of the phase diagram,
dynamical fluctuations grow with time
until they cause the decomposition of the nuclear system.
To be more definite, we fix the onset
of the phase separation at the instant when the
maximum of the variance $\sigma^2_k(t)$ becomes twice its
initial value for $k=0$.
In addition, we identify the probability $Y(A,t)$ with the size
distribution of the domains containing the liquid phase.
Our choice of the instant for the onset of
the phase separation is also supported by the observation
that this time roughly coincides with the beginning
of the plateau observed in the behavior of the ratio $L(t)/r_0(t)$.
Thus, while fluctuations continue to grow with time, the size
distribution $Y(A,t)$ does not change considerably.
The times for the beginning of the plateau in the three cases shown in
Fig.~\ref{fig3}, corresponding to
$\alpha=(1,1/3,1/6)10^{-2}~c/{\rm fm}$, are $t\sim 50,80,100\:{\rm fm}/c$
respectively.
\par
In order to assess the validity of our approach,
we compare the results of our calculations with the corresponding
experimental data by identifying the probability $Y(A,t)$ with
the distribution of the fragment yield observed in processes
of nuclear multifragmentation. The probability $Y(A,t)$ is
calculated with the value of $L(t)/r_0(t)$ at the
plateau. Like in Ref. \cite{Mat00}, finite--size and surface effects
are not included in our model, consequently a
transition from surface to bulk multifragmentation,
as suggested recently
by the analysis of the ISiS results \cite{Beau00},
cannot be described within our approach. However, we  expect
that our treatment should reliably account for the gross features
of the bulk instabilities of nuclear matter, at least at a
qualitative level.
\par
Since experimentally the fragments are detected
according to their charge, we have to transform  $Y(A,t)$
into the corresponding function of $Z$, $Y(Z,t)$. For this purpose, we
assume a
homogeneous distribution for the charge, $Z=[(1-I)/2]A$,
with $I=(N-Z)/A$ and use $I=0.2$, that corresponds
to the average asymmetry of the nuclear systems considered.
\par
\begin{figure}
\includegraphics{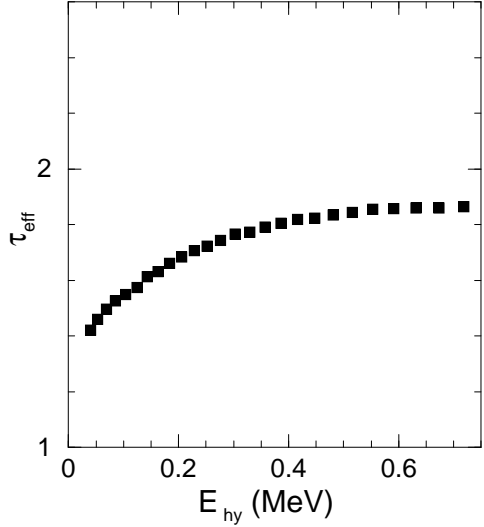}
\caption{\label{fig4}Effective exponent of the power law,  
$Y(Z)=\,Y_0Z^{-\tau_{eff}}$, 
as a function of the collective radial energy 
per particle at the break--up (the corresponding range of the expansion 
parameter is $(1/6)10^{-2}\leq \alpha \leq 10^{-2}$). 
}
\end{figure}

In Ref. \cite{Mat00} it has been shown that the probability $Y(Z)$
can be fitted with good accuracy by a power law
$Y(Z)=\,Y_0Z^{-\tau_{eff}}$. The parameter $\tau_{eff}$
depends only on the expansion rate of nuclear matter,
once the temperature and density at $t=0$ have been fixed. 
We have taken the initial values 
$T(0)=6\,{\rm MeV}$ and $\varrho_0(0)=0.7\varrho_{eq}$, however,
even if the temperature is changed in a rather large interval,
the value of the ratio $L(t)/r_0(t)$ 
at the plateau does not change appreciably. 
In Fig.~\ref{fig4} the effective exponent $\tau_{eff}$
is shown as a function of the radial collective
energy per particle, $E_{hy}$, at the break--up
of the nuclear system. This energy has a one--to--one
correspondence with the parameter $\alpha$, 
for given initial conditions, and is a more suitable observable 
than $\alpha$ for a comparison with the experimental data. 
Figure \ref{fig4} shows a moderate rise of $\tau_{eff}$ when  the
energy of the hydrodynamic motion increases. The cause of the expansion
can be either dynamical, following an initial compression, or thermal,
due to a release of energy in the bulk of the nuclear system
from very energetic particles, for instance
protons, antiprotons or pions. In both the cases the 
rapidity of the expansion and, as a
consequence, the exponent $\tau_{eff}$ increase with the
excitation energy of the nuclear system. 
\begin{figure}
\includegraphics{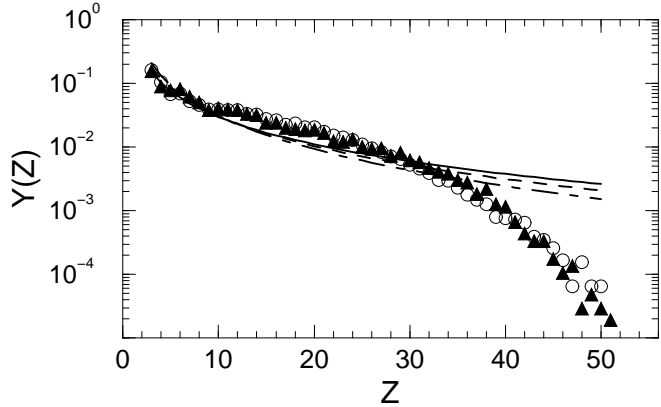}
\caption{\label{fig5}Comparison of fragment distribution
$Y(Z)$ calculated for $L(t)/r_0(t)=1.3$ (~dot--dashed line~) $1.6$, 
(~dashed line~) and $2$ (~solid line~) 
(~$\tau_{eff}=1.65,1.51,1.41$ respectively~), 
with experimental distributions for the reactions $^{129}Xe+Sn$ at
$E=32~A\,{\rm MeV}$ (triangles) and
$^{155}Gd+U$ at $E=36~A\,{\rm MeV}$ (circles) \cite{Fran01}. 
Both data and curves have been normalized to 1.} 
\end{figure}
\begin{figure}
\includegraphics{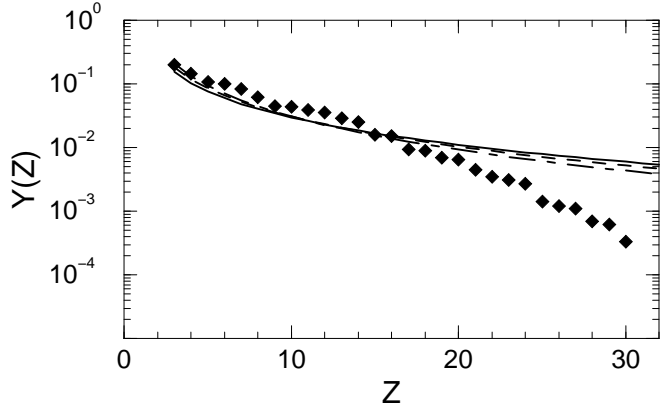}
\caption{\label{fig6}Same of Fig.~\ref{fig5}, but for the reaction $Xe+Sn$ 
at $E=50~A\,{\rm MeV}$ (diamonds)  \cite{Ass99}. 
}
\end{figure}
\par
In Fig.~\ref{fig5} a comparison is made between the charge
distributions predicted by our approach and  recent
experimental data obtained by the the INDRA Collaboration
for $^{129}Xe+Sn$ and
$^{155}Gd+U$ collisions at $E=32~A\,{\rm MeV}$ and at
$E=36~A\,{\rm MeV}$ respectively \cite{Fran01}.  In Fig.~\ref{fig6}, 
the same comparison is made for
$Xe+Sn$ collisions at $E=50~A\,{\rm MeV}$ \cite{Ass99}.
Our calculations have been performed for three different values of
the parameter $L(t)/r_0(t)$, the corresponding values
of the effective exponent $\tau_{eff}$ are given in the  legend of
Fig.~\ref{fig5}.
We have normalized
both the experimental and the calculated distributions
to one in order to perform the comparison on an absolute scale.
The agreement between experimental
data and our calculated charge distributions is quite
satisfactory for $Z<30\div 35$ at the lower incident energies of
Fig.~\ref{fig5},
and for $Z<20\div 24$ at the higher energy of Fig.~\ref{fig6}.
For the higher values of $Z$ the observed distributions
display a steeper slope than that predicted
by our calculations. This faster decrease of the experimental
yields should be
ascribed to finite--size effects \cite{Dag96} that have not
been included in our nuclear matter treatment.
The comparison also show that experimental data at higher
energy are better reproduced with smaller values of
the parameter $L(t)/r_0(t)$ and  correspondingly larger values
of $\tau_{eff}$, compared with the values needed 
to reproduce the data at lower energy. 
This trend is explained by our scheme, since a  growth
of the radial collective energy corresponds to an
increase of $\tau_{eff}$ ( see Fig.~\ref{fig4} ) and
it is reasonable to expect that for central collisions the
radial collective motion  becomes faster when
the incident energy increases.
\par
A further remark about the effective exponent $\tau_{eff}$.
It has  been  experimentally observed that
the onset of an expansion in nuclear matter is a signal
for a transition from surface--dominated to bulk
emission, as expected for a spinodal decomposition, and that
the power--law parameter $\tau_{eff}$ grows with
excitation energy for expanding nuclear matter
\cite{Beau99,Beau00}. Our results are in qualitative
agreement with this observed behavior and so is our estimate
of the radial collective energy. However, the experimental
values of $\tau_{eff}$ reported in Ref. \cite{Beau99}
are systematically larger than our values.

\section{Summary and Conclusions}

We have studied the density fluctuations of expanding nuclear matter within a
one--body treatment of nuclear dynamics. In our approach the
fluctuations are generated by adding a stochastic term to the
mean field and they develop about a local--equlibrium state
described by a hydrodynamic expansion.
This additional random force is constrained by an  extension of the usual
fluctuation--dissipation theorem to the case of uniformly
expanding systems. In this more general context, 
we have confirmed the main results obtained in  Ref. \cite{Mat00} 
about the particular physical conditions in which
a white--noise assumption for the stochastic field can be retained:
the density and temperature of the system should be
sufficiently close to the borders of the spinodal region
in the ($\varrho,T$) plane, so that the limit $\omega/T\ll 1$
gives a reasonable approximation to the density--density response.
In this case the equilibrium fluctuations can be adequately
described  by means of thermodynamic functions.

We have extended the results obtained for the
probability distribution of fluctuations in a system at local-equilibrium
to unstable situations. This has been done by extrapolating the relevant
quantities across the boundary of the spinodal region. Because
of the linear approximation used for evaluating the response
of the system to the stochastic force, the probability distribution
of the fluctuations is Gaussian. The variance of the distribution
at a given point in the ($\varrho,T$) plane, strongly depends
on the expansion velocity and instabilities are favoured in the case
of slower expansions.
\par
In Sec.~\ref{pp} we have discussed a procedure
to determine the size distribution of the domains
containing correlated density fluctuations. This procedure
is quite general and can be applied to any Gaussian
distribution. For the size distribution we have found an
expression similar to that obtained in Ref. \cite{Mat00}.
It again contains only one parameter, the ratio between the
dynamical correlation length and the mean interparticle
spacing, $L(t)/r_0(t)$. However, in the present case the time
behavior of $L(t)$ is mainly determined by the expansion
velocity, in other words by the collective energy stored in
the excited nuclear system. In particular, the parameter $L(t)/r_0(t)$
displays a flat behavior in time after a certain instant that depends
on the expansion velocity. Roughly at this time, the
amplitude of fluctuations acquires at least twice
its initial value and keeps growing. Therefore, the
decomposition of the system can be viewed as a quasi
stationary process.
\par
In Sec.~\ref{BB} we have compared the obtained mass distribution to the 
yield of intermediate--mass fragments observed in the multifragmentation
of heavy nuclei. The comparison with experiment has shown that our
approach fairly reproduces the measured charge distributions with
$Z<30\div35$ for  collisions at lower energy  and with $Z\lesssim 20$ at
higher energies. Since we have modeled the system as infinite nuclear matter,
we expect to overestimate the number of large fragments.
However, it could be worthwhile studying whether and how
the finite size effects could become more important with increasing
incident energy.
\par
Our approach does account for both the power--law distribution found
experimentally and the observed behavior of the effective exponent with the
excitation energy, but the calculated exponent is 
systematically smaller than that determined from experiments.
This discrepancy deserves further investigation.
\appendix*
\section{}
In this Appendix we briefly discuss the statistical properties
of nuclear matter that undergoes the expansion
described in Sec. \ref{oo}.
For this hydrodynamic motion
we can assume that in the local reference frame moving with
the hydrodynamic velocity ${\bf u}({\bf r},t)$, the
physical system is described by the equilibrium density matrix:
\begin{equation}
{\widehat \varrho}(t)=\frac{1}{Z(t)}e^{\displaystyle -\beta(t)\tilde H}\,,
\label{stat1}
\end{equation} 
where $Z(t)=Tr{\widehat \varrho}(t)$ is the partition function 
and $\tilde H$ is the Hamiltonian in the local reference frame   
\begin{equation}
\tilde H=\,\sum_j\frac{({\hat {\bf p}}_j-m{\bf u}({\bf r}_j,t))^2}
{2m}+\frac{1}{2}\sum_{i\not= j}V({\bf r}_i,{\bf r}_j).
\label{hamil}
\end{equation}
It can be shown that, for a generic couple of states $|m>$ and $|n>$ 
\[
<m|\tilde H|n>=<m|\,e^{ i\Sigma_j\chi({\bf r}_j,t)}\,
H\,e^{-i\Sigma_j\chi({\bf r}_j,t)}|n>\,,\]
where
$\chi({\bf r},t)$ is a gauge function such that
$\nabla\chi({\bf r},t)=m{\bf u}({\bf r},t)$. Thus,
the statistical density matrix of Eq. (\ref{stat1}) is
equivalent to
\begin{equation}
{\widehat \varrho}(t)=\frac{1}{Z(t)}\,
e^{ i\Sigma_j\chi({\bf r}_j,t)}
e^{\displaystyle -\beta(t) H} 
e^{-i\Sigma_j\chi({\bf r}_j,t)}\,.
\label{stat}
\end{equation}
Because of the properties of the trace the partition function 
\[Z(t)=Tr\,e^{\displaystyle -\beta(t)H}\]
is independent of the velocity field. The density matrix 
of Eq. (\ref{stat}) reproduces the correct expression  
both for the collective current and for the density of the   
hydrodynamic kinetic energy 
\[\frac{1}{2}<\sum_j\bigg(\frac{{\bf p}_j}{m}\,\delta({\bf r-r}_j)+
\delta({\bf r-r}_j)\,\frac{{\bf p}_j}{m}\bigg)>
=\,\varrho_0(t){\bf u}({\bf r},t)\]
and
\[<\sum_j\bigg(\frac{{\bf p}{_j}^2}{2m}\,\delta({\bf r-r}_j)\bigg)>
=\,\frac{1}{2}\varrho_0(t){\bf u}^2({\bf r},t)+{\cal E}_{kin}\,,\]
where ${\cal E}_{kin}$ is the density of the intrinsic kinetic energy. 
\par
The observables that commutate with ${\bf r}$
take the same mean value as for a homogeneous medium at equilibrium,
having the specific values of density and temperature at a
given instant. The same result holds for the equal--time correlation
functions of such observables. In particular, for the density
fluctuations we obtain
\begin{equation}
<\delta\varrho({\bf r},t)\delta\varrho({\bf r}^\prime,t)>
\,=\frac{1}{Z(t)}Tr\,\delta\varrho({\bf r})\delta\varrho({\bf r}^\prime)
e^{\displaystyle -\beta(t)H}\,.
\label{eqtfluct}
\end{equation} 
\par
As far as the correlation functions at different times are concerned, 
the result of Eq. (\ref{resp}) of Sec. \ref{oo} suggests that, 
when the collective expansion is slow with respect to the
Fermi velocity, these
functions depend only on the values of density and
temperature of the system at a given instant. In this case, we can
reasonably neglect the velocity
field $\sum_j\chi({\bf r}_j,t)$ in the expression of the
density matrix when evaluating the density correlations. Then
we expect that a instantaneous
fluctuation--dissipation relation still holds for a medium
undergoing a hydrodynamic uniform expansion. In fact, it
can be shown that
\begin{equation}
<(\delta\varrho_{\bf k}\delta\varrho_{-\bf k})(\omega)>_{t^\prime}=\,
-\frac{2}{1-e^{-\beta(t^\prime)\omega}}{\rm Im}\,D_k(\omega,t^\prime)\, ,
\label{fdr}
\end{equation}
where the time Fourier transforms have been evaluated respect to 
$t-t^\prime$. The statistical average is calculated with 
the density matrix 
\[{\hat \varrho}(t^\prime)=\frac{1}{Z(t^\prime)}
e^{-\beta(t^\prime)H}\,.\]

\end{document}